\documentclass{article}
\usepackage{amssymb}

\begin{document}

\title{On topological bias of discrete sources in the gas of wormholes }
\author{A.A. Kirillov, E.P. Savelova, G.D. Shamshutdinova \\
%EndAName
\emph{Branch of Uljanovsk State University in Dimitrovgrad, }\\
\emph{Dimitrova str 4.,} \emph{Dimitrovgrad, 433507, Russia} }
\date{}
\maketitle

\begin{abstract}
The model of space in the form of a static gas of wormholes is considered.
It is shown that the scattering on such a gas gives rise to the formation of
a specific diffuse halo around every discrete source. Properties of the halo
are determined by the distribution of wormholes in space and the halo has to
be correlated with the distribution of dark matter. This allows to explain
the absence of dark matter in intergalactic gas clouds. Numerical estimates
for parameters of the gas of wormholes are also obtained.
\end{abstract}

\section{Introduction}

As it was shown recently all the variety of dark matter phenomena can be
prescribed to the locally non-trivial topological structure of space or
equivalently to the existence of the specific topological bias for sources
(e.g., see Ref. \cite{K06}). The simplest model of such a space is given by
a static gas of wormholes which was shown to give rise to the scale
dependent renormalization of intensities of pure gravitational sources,
i.e., to the origin of "dark matter halo" \ around any point-like
gravitating mas \cite{KS07}. In the present paper we show that together with
the effect pointed above out the gas of wormholes gives rise to an analogous
renormalization of all cosmic discrete sources of radiation. By other words
any discrete source turns out to be surrounded with a diffuse halo.
Moreover, by virtue of their common origin, such a halo should strongly
correlate with the dark matter halo. The basic results of the present paper
are the expressions \ (\ref{bw}) and (\ref{bbr}) for the halo and numerical
estimates for the astrophysical parameters of the gas of wormholes (mean
density, the characteristic scale of the throat, relative brightness of the
halo). When observing galaxies such a diffuse halo has a low surface
brightness and is usually considered as a cosmic background which has
different from radiation of the galaxy origin. We recall that usually, the
observed diffuse halos in galaxies are attributed to reflection from dust,
and the general diffuse component is assumed to originate from very fade and
remote galaxies. This leads to an essential overestimation of the
mas-to-luminosity ratio in galaxies $M/\ell $ and, therefore, this is
usually interpreted as the presence of dark matter. As we show in the
present paper such a ratio should be bigger $M/\ell \gg 1$ in smaller
objects (e.g. dwarfs galaxies), while in larger objects (e.g., cluster size
plasma clouds), due to their huge size (more than or of the order of the
characteristic scale of the halo), radiation from the halo sums always up
with radiation from the hot cloud itself and the parameter $M/\ell \sim 1$,
e.g., dark matter is always absent \cite{DMpaper}.

We recall that nontrivial topological structure (or the topological bias)
was to form during quantum stage of the evolution of the Universe, when
spacetime topology underwent fluctuations and the spacetime itself had the
foam-like structure \cite{wheeler}. During the expansion the Universe cools
down, quantum gravity processes stop and the topological structure tempers.
There are no convincing theoretical arguments of why such a foam-like
structure of space should decay upon the quantum period. Moreover, the
presence of a considerable portion of dark energy (i.e. of an effective
cosmological constant) in the present Universe and in the past, on the
inflationary stage \cite{inf}, may be considered as the very basic
indication of a nontrivial topological structure of space \cite{KS08}

Indeed, dark energy violates the weak energy condition $\varepsilon +3p>0$.
Save speculative theories (or pure phenomenological models \cite{inf}),
there is no matter which meets such a property. However in the presence of a
non-trivial topology, vacuum polarization effects are known to give rise
quite naturally to such a form of matter \cite{Grib}. By other words, up to
date the only rigorous way to introduce dark energy is to consider the
vacuum polarization effects on manifolds of a non-trivial topological
structure. It is necessary to point out that stability of wormholes requires
the presence of matter violating the weak energy condition. In particular,
wormholes are known to be possibly supported by the vacuum polarization
induced by the wormholes themselves (e.g., see Ref. \cite{whst}).

\section{The topological bias of sources}

Consider a unite discrete source of radiation and the problem of scattering
on the gas of wormholes. For the sake of simplicity we consider the case of
a static gas, i.e., we assume that wormholes do not move in space. In this
case the scattering is not accompanied with the frequency shift. We point
also out that our results can easily be generalized to the case of the
expanding Universe.

The problem of the modification of the Newton's law in the presence of the
gas of wormholes (i.e., origin of  the topological bias) was solved recently
in Ref. \cite{KS07}. It turns out that in the case of radiation we can also
speak of the topological bias of sources. We shall be interested in \ the
behavior of the bias on scales $L\gg a$, where $a$ is a characteristic size
of a wormhole throat and, therefore, in such an approximation the throat
looks like a point object. Our aim is to find the Green function to the wave
equation
\begin{equation}
\left( k^{2}+\nabla ^{2}\right) G(r,r_{0})=4\pi \delta (r-r_{0})  \label{gg}
\end{equation}%
in the presence of the wormhole. Recall that that equation describes the
distribution of radiation on the frequency $\omega =kc$ (i.e., any component
$E$, $H$ or the vector potential $A)$ produced by a unite stationary source $%
j\sim e^{-i\omega }\delta (r-r_{0})$. Indeed, in the Lorentz gauge $\partial
_{i}A^{i}=0$ the Maxwell equations take the form%
\begin{equation}
\frac{\partial ^{2}}{\partial x^{k}x_{k}}A^{i}=\frac{4\pi }{c}j^{i},
\end{equation}%
where $j^{i}$ is $4$--current. Using now the linear relation between the
strength tensor and the vector potential $F_{ik}=\partial _{i}A_{k}-\partial
_{k}A_{i}$ we see that both $E_{i}=F_{0i}$ and $H_{i}=\frac{1}{2}\epsilon
_{ijk}F_{jk}$ obey the above equation with the obvious replacement $j^{i}$ $%
\rightarrow $ $J_{ik}=\partial _{i}j_{k}-\partial _{k}j_{i}$. Note that in
the case of an arbitrary source the vector potential (and respectively field
strengths) can be expressed via the Green function as follows%
\begin{equation}
A^{k}=\frac{1}{c}\int G(r,r_{0})j^{k}(r_{0})d^{3}r_{0}.
\end{equation}%
In the case of the flat space the (retarded) Green function is known to have
the standard form $G_{0}(R)=e^{ikR}/R$ (where $R=\left\vert
r-r_{0}\right\vert $). Due to the conformal invariance of the Maxwell
equations, the same function can be used for the class of conformally flat
metrics.

The simplest wormhole can be constructed as follows. Consider two spheres $%
S_{\pm }$ of the radius $a$ and at the distance $d=\left\vert
R_{+}-R_{-}\right\vert $ between their centers. The interior of the spheres
is removed and the surfaces of the spheres are glued together. Such spheres $%
S_{\pm }$ can be considered as conjugated mirrors, so that while the
incident signal fells on one mirror the reflected signal outgoes from the
conjugated mirror. Thus, every wormhole is determined with a set of
parameters $a$, $\mathbf{{R_{\pm }}}$, and $U$, where $a$ is the radius of
the throat, $\mathbf{{R_{\pm }}}$ stands for positions of centers of spheres
(i.e. of throats), and $U$ stands for the rotation matrix which defines the
gluing procedure for the surfaces of the spheres. In the approximation used
below the dependence on $U$ disappears and will not be accounted for.

The exact solution of the scattering problem in the case of a unique
wormhole is rather tedious and will be presented elsewhere. For
astrophysical needs it is sufficient to consider diffraction effects in the
geometrical optics limit $ka\gg 1$. According to the Huygens principle the
scattering on a wormhole can be prescribed to the presence of secondary
sources on throats which can be accounted for by additional terms
\begin{equation}
G(R)=G_{0}\left( R\right) +u_{R}^{+}-u_{A}^{+}+u_{R}^{-}-u_{A}^{-},
\end{equation}%
where terms $u_{A,R}^{\pm }$ describe absorption and reflection by throats $%
S_{\pm }$ respectively. Every such term can be described by the surface
integral (e.g., see the standard book \cite{BW})
\begin{equation}
u_{\alpha }^{\pm }\left( r,r_{0}\right) =\frac{k}{2\pi i}\int\limits_{S_{\pm
}^{\alpha }}G_{0}(r^{\prime },r_{0})\frac{e^{ikR^{\prime }}}{R^{\prime }}%
df_{n}
\end{equation}%
where $R^{\prime }=\left\vert r^{\prime }-r\right\vert $, and $S_{\pm }^{R,A}
$ denotes the lighted and dark sides of throats respectively\footnote{%
We note that the lighted side of a throat $S_{\pm }$ is turned by the matrix
$U^{\pm 1}$ with respect to the dark side, so that in general the union $%
S^{R}\cup S^{A}\neq S$.}. If we neglect the throat size then the integration
gives the square $\pi a^{2}$. Then we find terms which describe reflection
and absorption of the signal and correspond to secondary sources placed on
the throats
\begin{equation}
u_{A}^{\pm }=\frac{k}{2\pi i}\pi a^{2}G_{0}(R_{\pm }-r_{0})G_{0}(r-R_{\pm }),
\end{equation}%
\begin{equation}
\ u_{R}^{\pm }=\frac{k}{2\pi i}\pi a^{2}G_{0}(R_{\mp }-r_{0})G_{0}(r-R_{\pm
}),
\end{equation}%
which are spherical waves from the two additional sources at the positions $%
\vec{R}_{\pm }$. In this manner we see that the scattering on wormholes can
be prescribed to additional sources, i.e., to the bias of the point source
in (\ref{gg}) of the form%
\begin{equation}
\delta (r-r_{0})\rightarrow \delta (r-r_{0})+b\left( r,r_{0}\right) ,
\label{bf}
\end{equation}%
where $b\left( r,r_{0}\right) $ is the bias function. Indeed, in this case
the Green function remains formally the same as in the flat space $G_{0}(R)$%
, while the scattering on the topology is described by the bias of sources $%
J(r)\rightarrow J(r)+\int b(r,r^{\prime })J(r^{\prime })d^{3}r^{\prime }$.
Thus, in the case of the static gas of wormholes the bias function takes the
form
\begin{equation}
b\left( r,\omega \right) =\frac{\omega }{2\pi ic}\sum\limits_{m}\pi
a_{m}^{2}\left( \frac{e^{ikR_{-}^{m}}}{R_{-}^{m}}-\frac{e^{ikR_{+}^{m}}}{%
R_{+}^{m}}\right) \left[ \delta (\vec{r}-\vec{R}_{+}^{m})-\delta (\vec{r}-%
\vec{R}_{-}^{m})\right] ,  \label{bw}
\end{equation}%
where for the sake of simplicity we set $r_{0}=0$, and the index $m$
enumerates different wormholes.

Thus the presence of the gas of wormholes leads to the origin of a specific
radiating halo (\ref{bw}) around every point source. Due to the randomness
of phases in multipliers in (\ref{bw}) such a halo has incoherent (or
diffuse) nature.

Using the density of distribution of wormholes $F(R_{-},R_{+},a)$ and
transforming sums in integrals the above equation can be cast to the form%
\begin{equation}
b\left( r,\omega \right) =\frac{\omega }{2ic}n\int \left( \frac{e^{ikR}}{R}-%
\frac{e^{ikr}}{r}\right) \left[ g\left( R,r\right) +g\left( r,R\right) %
\right] d^{3}R
\end{equation}%
where $g=\frac{1}{n}\int a^{2}F\left( R_{-},R_{+},a\right) da$ and $n$
denotes the mean density of wormholes in space. In the case of a homogeneous
distribution of wormholes the function $g$ depends only on $d=|R_{+}-R_{-}|$%
. Then the bias function takes the simplest form for the Fourier transforms
\begin{equation}
b\left( k,\omega \right) =\frac{\omega }{ic}n\frac{4\pi \left( g(k)-g\left(
0\right) \right) }{k^{2}-\frac{1}{c^{2}}(\omega +io)^{2}}.  \label{br}
\end{equation}%
where $g(k)=(2\pi )^{-3/2}\int g(r)e^{-ikr}d^{3}r$ is the Fourier transform
for the function $g(\vec{d})$.

We note that since we are working in the geometric optics limit, the Lorentz
invariance and the standard dispersion relations for photons are not
violated. Nevertheless, the violations surely take place at the scales,
where the geometric optics does not work. Namely, at sufficiently large
distances where topological defects start to show up (e.g., dark matter
starts to show up at $\lambda =2\pi /k\gtrsim $ of the order of a few $Kpc$
\cite{K06} which is much more, than any wavelength $\lambda \sim c/\omega $
of a photon detected).

The violation of the Lorentz invariance has intensively been discussed in
the literature \cite{Aether}. All the existing estimates coincide by the
order of magnitude and concern only of extremely small scales \cite{K,sr}.
In particular in a recent paper \cite{sr} it was established the well-known
(e.g., see also Refs. \cite{K,l}) very strong limit for the "first"
correction to the standard dispersion relation. Namely, if $\omega
^{2}=k^{2}(1+kl_{1}+k^{2}l_{2}^{2}+...)$, then $l_{1}<L_{Pl}$, where $L_{Pl}$
is the plankian length. \ We point out that such a correction corresponds to
the decomposition over the small parameter $kl\ll 1$, where $l$ has the
sense of a characteristic scale connected to wormholes\footnote{%
Recall that wormholes relate to the three basic parameters. Those are the
density of wormholes $n^{-1/3}$, the mean throat size $a$, and the mean
distance between throats $d=\left\vert R_{+}-R_{-}\right\vert $.}, which
leads to a very strong limit on the existence of microscopic wormholes \
(i.e., we admit only the existence of wormholes at scales $l_{1}<L_{Pl}$,
see also analogous estimates in Refs. \cite{K,l}).

However, (we hope that this should not be a surprise for readers) the same
restriction can be used to estimate parameters of wormholes having the
astrophysical meaning (scales). Indeed, in the presence of wormholes of
astronomical scales ($L\sim $ a few $Kpc$), the small parameter is already
the ratio of the photon wavelength to the characteristic scale of a
wormhole, which is the inverse parameter $1/(kL)$. Therefore, in the
dispersion relation the first non-trivial corrections should have the form $%
\omega ^{2}=k^{2}\left( 1+\frac{1}{kL_{1}}+\frac{1}{k^{2}L_{2}^{2}}%
+...\right) $. Then, if we use the above restrictions on the absolute value
of the existing correction to the dispersion relation (e.g., $k\sim
10^{6}cm^{-1}$ and  $kl_{1}<10^{-27}$), then we find that the characteristic
scales of astrophysically significant wormholes should obey the
"restrictions" $L_{1}>(1\div 5)Kpc$. Those are just the scales at which dark
matter effects start to display themselves and according to our
interpretation of dark matter effects \cite{K06} the density of wormholes
should achieve the value $nL_{1}^{3}\sim 1$. We also point out to the fact
that at galaxy scales the Lorentz invariance violates also due  to
relativistic gravitational effects (i.e., due to general relativity). By
other words to essentially improve the above restrictions on the dispersion
relations violation is hardly possible.

As it was pointed out above the reason of the absence of the Lorentz
invariance violation, e.g., in the optic range, is very simple. For rather
short wave-length the forming halo secondary sources (cf. expression (\ref%
{bw})) have random phases and, therefore, do not contribute to the amplitude
of a signal. By other words the halo carries the diffuse character. In such
a case it is more correct to consider the expression for the energy flux
which comes to the point $r$ which for the diffuse field becomes an additive
quantity.

\section{The diffuse halo}

Consider now the renormalization of the intensity of radiation. By virtue of
the diffuse character of the current the intensity of radiation is
determined by the square of the current, e.g.,
\begin{equation}
\overline{I(R)I^{\ast }(R^{\prime })}=\left\vert I(R)\right\vert ^{2}\delta
(R-R^{\prime }),
\end{equation}%
where the averaging out should be thought as either over the period of the
field $T=2\pi /\omega $, or over the random phases. When we do not account
for the scattering on topology, i.e., in the ordinary flat space, the
intensity of radiation $W=\frac{E^{2}+H^{2}}{8\pi }$ is determined by the
intensity of the current $|I|^{2}$ as $W(r)\sim \int \frac{|I(r)|^{2}}{%
|r-r^{\prime }|^{2}}d^{3}r^{\prime }$ and, in particular, for a point source
$|I(r)|^{2}=|I_{0}|^{2}\delta (r-r_{0})$ the intensity of the energy flow is
$W(r)=|G_{0}(r-r_{0})|^{2}{|I_{0}|^{2}}$. The scattering on wormholes leads
to the replacement $G_{0}\rightarrow G$ which effectively can be described
as the origin of the bias (\ref{bf}), (\ref{bw}) or as the origin of an
additional halo which has the property pointed above out to be
delta-correlated and which leads \ to a renormalization of the current
intensity $|I_{0}|^{2}\rightarrow |\tilde{I}(r)|^{2}$. We shall use the
Green function, i.e., we consider the sum
\begin{equation}
G(r)G^{\ast }(r)=\left\vert G(r)\right\vert ^{2}=\left\vert G_{0}\left(
r\right) \right\vert ^{2}+\sum_{s=A,R;\ p=\pm }\left\vert
u_{s}^{p}\right\vert ^{2}
\end{equation}%
which gives
\begin{equation}
\left\vert G(r)\right\vert ^{2}=\frac{1}{r^{2}}+\frac{\omega ^{2}}{4\pi
^{2}c^{2}}\sum\limits_{m}\pi ^{2}a_{m}^{4}\left( \frac{1}{\left(
R_{-}^{m}\right) ^{2}}+\frac{1}{\left( R_{+}^{m}\right) ^{2}}\right) \left(
\frac{1}{\left\vert R_{-}^{m}-r\right\vert ^{2}}+\frac{1}{\left\vert
R_{+}^{m}-r\right\vert ^{2}}\right) .
\end{equation}%
In the above expression due to the randomness of phases the intersection
terms are omitted. Then the above expression defines the bias for the
intensity of a unite source in the form
\begin{equation}
\left\vert G(r)\right\vert ^{2}=\frac{1}{r^{2}}+\int \frac{\tilde{b}^{2}(R)}{%
\left\vert R-r\right\vert ^{2}}d^{3}R
\end{equation}%
where
\begin{equation}
\tilde{b}^{2}(R)=\frac{\omega ^{2}}{c^{2}}\frac{n}{4}\int \left( \frac{1}{%
R^{2}}+\frac{1}{X^{2}}\right) \left( \tilde{g}\left( R,X\right) +\tilde{g}%
\left( X,R\right) \right) d^{3}X
\end{equation}%
and $\tilde{g}\left( R_{+},R_{-}\right) =\frac{1}{n}\int a^{4}F\left(
R_{-},R_{+},a\right) da.$ For an isotropic distribution of wormholes $\tilde{%
g}=\tilde{g}\left( \left\vert R_{+}-R_{-}\right\vert \right) $ and therefore
we find the bias function in the form%
\begin{equation}
\tilde{b}^{2}(R)=\frac{k^{2}}{2}n\int \left( \frac{1}{R^{2}}+\frac{1}{%
\left\vert X+R\right\vert ^{2}}\right) \tilde{g}\left( X\right) d^{3}X.
\label{bbr}
\end{equation}

In this manner the relation between the intensity of an actual $%
|I_{0}(r)|^{2}$ and the apparent $|\tilde{I}(r)|^{2}$ currents (or, which is
equivalent, between the actual $\ell _{0}$ and observed $\ell $ luminosity)
is determined by the distribution of wormholes in space and is given by the
expression
\begin{equation}
|\tilde{I}|^{2}(r)=|I_{0}|^{2}(r)+\int \tilde{b}^{2}(r-r^{\prime
})|I_{0}|^{2}(r^{\prime })d^{3}r.
\end{equation}

\section{Estimates}

Consider now some simplest estimates. In order \ to find estimates for the
renormalization of the surface brightness of a source we consider the case
when all wormholes have the same value for $d=\left\vert \vec{R}_{-}-\vec{R}%
_{+}\right\vert =r_{0}$. In this case we can take $\tilde{g}\left( X\right) =%
\frac{\overline{a^{4}}}{4\pi r_{0}^{2}}\delta \left( X-r_{0}\right) $ and
for a point source we get the bias in the form
\begin{equation}
\tilde{b}^{2}(R)=\frac{k^{2}}{2}n\overline{a^{4}}\frac{1}{R^{2}}\left( 1+%
\frac{R}{2r_{0}}\ln \left\vert \frac{R+r_{0}}{R-r_{0}}\right\vert \right) .
\label{bbw}
\end{equation}%
We note that the characteristic behavior $\tilde{b}^{2}\sim 1/R^{2}$ of the
halo density is the specific attribute of the point-like structure of a
source. In the case of actual sources such a halo acquires the cored
character $\tilde{b}^{2}\sim \tilde{b}^{2}(l)\sim const$, where $l$
corresponds to the linear size of the source.

To get the estimate to the number density of wormholes is rather
straightforward. First wormholes appear at scales when dark matter effects
start to display themselves, i.e., at scales of the order $L_{1}\sim (1\div
5)Kpc$, which gives in that range \ the number density
\begin{equation}
n\sim (3\div 0.024)\times 10^{-66}cm^{-3}.  \label{n}
\end{equation}

The characteristic size of throats can be estimated as follows \cite%
{KS07,KS08}. As it was pointed out above in the case of a homogeneous
distribution of wormholes the value of $\bar{a}$ defines the amount of dark
energy in the Universe. Indeed, consider a single wormhole in the flat
(Minkowsky) space. Then the metric can be taken in the form (e.g., see Ref.
\cite{KS07})
\begin{equation}
ds^{2}=dt^{2}-f^{2}\left( r\right) (dr^{2}+r^{2}\sin ^{2}\vartheta d\phi
^{2}+r^{2}d\vartheta ^{2}),
\end{equation}%
where $f\left( r\right) =1+\theta (a-r)(\frac{a^{2}}{r^{2}}-1)$ and $\theta
(x)$ is the step function. We can replace $f\left( r\right) $ with any
smooth function, this however will not change the subsequent estimates. Both
regions $r>a$ and $r<a$ represent portions of the ordinary flat Minkowsky
space and therefore the curvature is $R_{i}^{k}\equiv 0$. However on the
boundary $r=a$ it has the singularity which defines the scalar curvature as $%
R=-8\pi GT=\frac{2}{a}\delta \left( r-a\right) $ where $T$ stands for the
trace of the stress energy tensor which one has to add to the Einstein
equations to support such a wormhole. It is clear that such a source
violates the weak energy condition and, therefore, it reproduces the form of
dark energy (i.e., $T=\varepsilon +3p<0$). If the density of such sources
(and respectively the density of wormholes) is sufficiently high, then this
results in an acceleration of the scale factor for the Friedmann space as $%
\sim t^{\alpha }$ with $\alpha =\frac{2\varepsilon }{3(\varepsilon +p)}>1$,
i.e., see Refs. \cite{inf}.

Every wormhole gives contribution $\int Tr^{2}dr$ $\sim $ $\bar{a}$ to the
dark energy, while the dark energy density is $\varepsilon _{DE}$ $\sim $ $%
(8\pi G)^{-1}n\bar{a}$. Since the density of dark energy has the order $%
\varepsilon _{DE}\sim $ $0.75\varepsilon _{0}$, where $\varepsilon _{0}$ is
the critical density, then we immediately find the estimate $\bar{a}\sim
(1\div 125)\times 10^{-3}R_{\odot }$, where $R_{\odot }$ is the Solar
radius. Now by means of use of the expression (\ref{bbw}) we find the
estimate for the relative brightness of the halo
\begin{equation}
\ell /\ell _{0}\sim \frac{k^{2}}{2}n\overline{a}^{4}l\sim 4\frac{l}{R_{\odot
}}\left( \frac{k}{k_{0}}\right) ^{2}\left[ 1\div 2\times 10^{6}\right]
\times 10^{-14}.
\end{equation}%
Here $l$ is the linear size of the source around which the diffuse halo
forms, and $k_{0}$ defines the wavelength $\lambda _{max}$ which corresponds
to the temperature $T_{\odot }=6\times 10^{3}K$. It is clear that the
relative brightness of the halo is small $\ell /\ell _{0}\ll 1$ and it
reaches the order of the unity only for sufficiently extended objects of the
characteristic size $\left[ 0.5\times 10^{-6}\div 1\right] \times
10^{14}R_{\odot }$. We also point out that outside the radiating region the
halo brightness decays according to (\ref{bbw}) as $\sim 1/R^{2}$.

\section{Acknowledgment}

This research was supported in part by RFBR 09-02-00237-a. We are also
grateful to A.A. Starobinsky for valuable comments.


\begin{thebibliography}{99}
\bibitem{K06} Kirillov, A.A., Phys. Lett. \textbf{B, 632}, 453 (2006);
Kirillov, A.A., Turaev, D., 2006, MNRAS, \textbf{371}, L31; Phys. Lett.
\textbf{B, 656}, 1 (2007).

\bibitem{KS07} Kirillov A.A., Savelova E.P. Phys. Lett. \textbf{B660}, 2008,
p. 93-99.

\bibitem{DMpaper} Clowe, D., et al., Astrophys. J. Lett.648 (2006) L109.

\bibitem{wheeler} Wheeler J.A. , (1964) in: \emph{Relativity, Groups, and
Topology, } B.S. and C.M. DeWitt (eds. ), Gordan and Breach, New York; S.W.
Hawking, Nuclear Phys., \textbf{B114} 349 (1978).

\bibitem{inf} A.A.Starobinsky, Phys. Lett. \textbf{B91}, 100 (1980);
A.H.Guth, Phys. Rev. \textbf{D23}, 347 (1981); A.A.Linde, Phys. Lett.
\textbf{B108}, 389 (1982).

\bibitem{KS08} Kirillov A.A., Savelova E.P., Gravitation and Cosmology Vol.
\textbf{14}, No3, pp. 256-261 (2008); Kirillov, A.A, Savelova E.P. Zolotarev
P.S., Phys. Lett. \textbf{B, 663}, 372-376 (2008).

\bibitem{Grib} A.A. Grib S.G. Mamaev V.M. Mostepanenko. Vacuum quantum
effects in intensive fields. M: Energoatomizdat, 1988 [in russian].

\bibitem{whst} V. Khatsymovsky, Phys. Lett. \textbf{B, 320}, 234-240 (1994).

%50

\bibitem{BW} L.D. Landau, E.M. Lifshitz. The field theory. Moscow: Nauka,
1973 [in russian]. %M. Born, E. Wolf, Principles of Optics, (Pergamon
%press, New York, 1968);
%\bibitem{jackson}

%Jackson J.D., Classical Electrodynamics, ed. (Wiley, New York,
%1962).

\bibitem{Aether} T. Jacobson, S. Liberati and D. Mattingly, Annals Phys.
\textbf{321}, 150 (2006); C. Eling, T. Jacobson and D. Mattingly,
\textquotedblleft Einstein-aether theory,\textquotedblright\
arXiv:gr-qc/0410001.

\bibitem{K} F.R. Klinkhamer, Nuclear Phys., \textbf{B578} 277 (2000); F.R.
Klinkhamer and C. Rupp, Phys. Rev. \textbf{D70}, 045020 (2004); Phys. Rev.
\textbf{D72}, 017901 (2005); S. Bernadotte, F.R. Klinkhamer, Phys. Rev.
\textbf{D 75}, 024028 (2007).

\bibitem{sr} A.A. Abdo et al. arXiv:0908.1832

\bibitem{l} M. Jankiewicz, R. V. Buniy, T. W. Kephart, T.J. Weiler,
Astropart. Phys \textbf{21, }651-666, (2004); W.A. Christiansen, Y. Jack Ng,
H. van Dam, Phys. Rev. Lett. \textbf{96}, 051301 (2006).
\end{thebibliography}
\end{document}